\documentclass[doublespacing]{elsart} 
 
\usepackage{graphicx} 
 
\usepackage{natbib} 
 
 \usepackage{epsfig} 
 
\usepackage{amssymb}

\begin{document} 
 
\begin{frontmatter}

\title{ Calibration of the mirror system in the HERA-B RICH}

 \author[JSI]{Marko Stari\v{c}\corauthref{cor1}} \ead{marko.staric@ijs.si}
 \corauth[cor1]{Corresponding author} and  \author[ULFMF,JSI]{Peter Kri\v{z}an}

 \address[JSI]{Jo\v{z}ef Stefan Institute, Ljubljana, Slovenia} 
 \address[ULFMF]{Faculty of Mathematics and Physics, University of Ljubljana, Slovenia}

\begin{abstract} 
The mirror system of the HERA-B RICH consists of two spherical and two planar mirrors, composed
of altogether 116 mirror segments. Analysis of displacements of the \v Cerenkov ring center
relative to the charged particle track, for given spherical-planar segment pairs, leads to
accurate information regarding the orientation of individual mirror segments. The method 
is described and the effect of applying the required corrections on the \v Cerenkov angle resolution
of the HERA-B RICH is discussed.
\end{abstract}

 
\end{frontmatter} 
 
\section{Introduction}  
 
\begin{figure}[hbt] 
\centerline{\epsfig{file=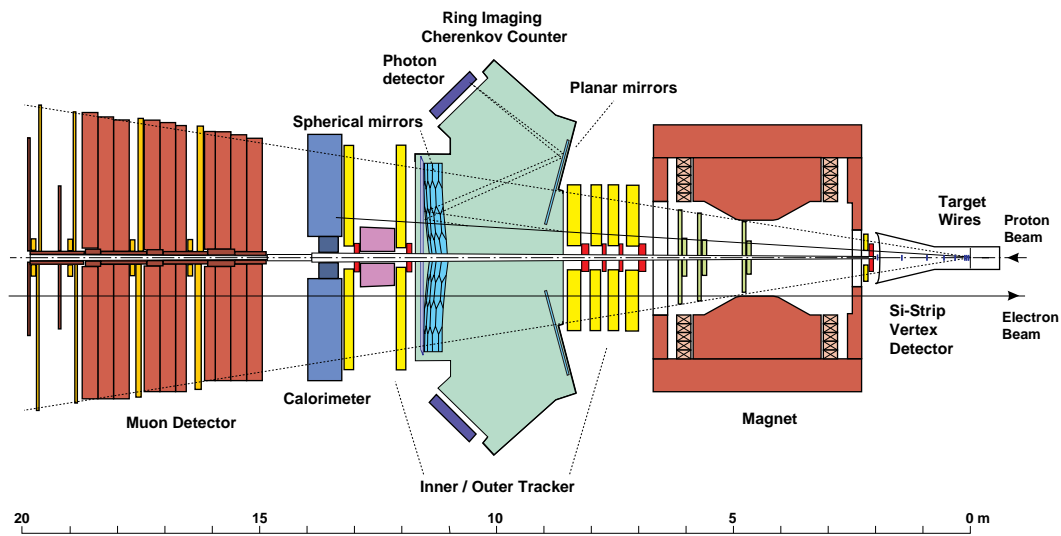,width=14.cm,clip}} 
\caption[kk]{  A side  view of the HERA-B detector.  
Photon paths in the RICH counter are indicated. 
} 
\label{hbdet} 
\end{figure}

HERA-B  \cite{prop} was  a  fixed  target experiment (Fig.~\ref{hbdet}) at  the HERA 
storage ring at DESY in Hamburg. The experiment used 920 GeV protons from the beam 
halo and a set of eight thin ribbons, of different materials, as targets. 
The interaction rate was adjusted by moving  
the targets in or out of the beam halo \cite{Target}. The experiment utilized a  
forward spectrometer capable of measuring interaction rates up to 40 MHz.  
The spectrometer consisted of a dipole magnet, 
a vertex detector \cite{VDS} upstream and 
a main tracking system downstream of the magnet \cite{ITR,OTR}. 
Particle identification was performed by a Ring Imaging \v Cerenkov (RICH) detector \cite{rich_nim}, 
an electromagnetic calorimeter \cite{ECAL} and a muon detector system \cite{MUON}. 
In addition, the experiment included a sophisticated hardware trigger 
for lepton track pairs to record leptonic decays of $J/\psi$ particles.  
The large acceptance of the spectrometer coupled with high-granularity particle 
identification devices and a precision vertex detector allowed for detailed studies  
of multi-particle final states 
\cite{PQ}-\cite{open-charm}. 
By using targets of different materials, HERA-B was 
also able to study the dependence of various properties of proton-nucleus interactions  
as a function of atomic number.

The identification of pions, kaons and protons was performed by the RICH detector 
\cite{rich_nim}. The HERA-B RICH  used atmospheric pressure $C_4F_{10}$ as \v Cerenkov 
radiator (n=1.00137). The focusing of \v Cerenkov light was achieved with two spherical 
mirrors,  tilted by  $9^0$ in  opposite  directions  
(Fig.~\ref{hbdet}).  Two 
planar mirrors then  reflected the light to photon  detectors at the top 
and bottom of the vessel  containing the radiator gas.  For the detection 
of \v Cerenkov photons multi anode PMTs (Hamamatsu  R5900) were used. 
The  inner  part of  the  photon  detector  surface was  equipped  with 
16-channel  PMTs (16 x 4 mm x 4 mm)  and  the  outer  region with  coarser granularity 
had 4-channel  PMTs (4 x 8 mm x 8 mm).  To overcome the loss of photons due to  
inactive space  between  PMT photocathodes, a  demagnifying lens 
system \cite{lens} was placed in front of each PMT.

The particle identification capabilities of a RICH counter are 
determined by the resolution of the measured  \v Cerenkov angle, 
which is given by the two main parameters of a RICH counter, the  
 \v Cerenkov angle 
resolution due to a single photon and the number of detected photons per \v Cerenkov ring. 
The measured average number of detected photons for particles approaching  
the speed of light amounts to 33. It is in good agreement with  
the value expected from  the data available on the quantum efficiency,  
mirror reflectivity, and transmissions of the vessel 
window and of the optical system \cite{rich_nim}.

The main contributions to the single photon resolution (r.m.s.) come  
from the photon detector granularity (0.50~mrad and 0.93~mrad for the  
regions covered by finer and coarser granularity PMTs respectively)  
and the dispersion in the radiator medium (0.33~mrad).   
The optical error (0.25~mrad) includes contributions from spherical 
aberration, 
mirror quality, and mirror alignment. The contribution of multiple  
scattering in the RICH counter ($\frac {3.5\;\rm{mrad}}{p\;\rm{(GeV/c)}}$)  
becomes important at low momenta. 
The resulting expected single photon resolution, 
$0.65\;\rm{mrad} \oplus \frac {3.5\;\rm{mrad}}{p\;\rm{(GeV/c)}}$ 
 and $1.02\;\rm{mrad} \oplus \frac {3.5\;\rm{mrad}}{p\;\rm{(GeV/c)}}$ 
for the regions covered by the two types of PMTs, does not include 
the contribution from the uncertainty in the track direction, which is given by 
other components of the HERA-B detector. 

In order to reach and maintain the optimal performance of the RICH counter, 
elaborate alignment and calibration methods have to be used. 
For the optical system of the HERA-B RICH, a calibration  method was 
used which is based on  a procedure originally developed on  simulated data \cite{align_mc}.

\section{Calibration of the optical system}

\begin{figure}[thb] 
\centerline{\hbox{\psfig{figure=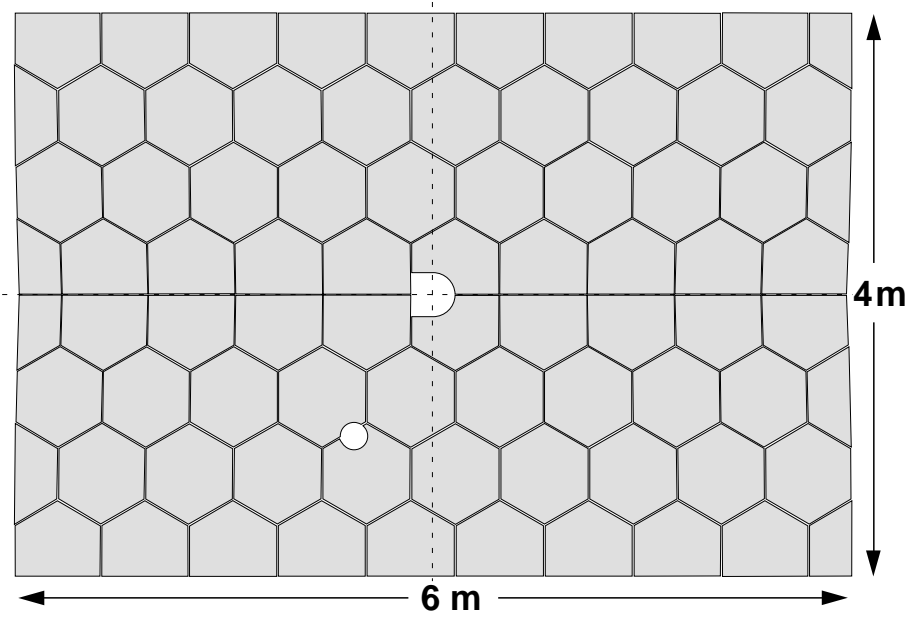,width=12cm}}} 
\caption[kk]{  Distribution of spherical mirror polygons. The holes in the 
          array are for the proton and electron beam pipes.} 
\label{mirror-tiling} 
\end{figure} 
 
The main imaging device of the HERA-B RICH  
is a spherical mirror placed inside the 
radiator vessel with the center of the sphere near the target and a radius of 
curvature of 11.4~m. The mirror, a 6~m by 4~m rectangular cutout from 
the sphere, consists of 80 full or partial hexagons (see 
Fig.~\ref{mirror-tiling}). To achieve a focal surface  
outside of the main particle 
flux ($\pm$160 mrad vertically), the mirror is split horizontally, and 
both halves are tilted by 9$^{\circ}$ away from the beam-line. A set 
of two planar mirrors, composed of 18 rectangular elements each, 
translates the focal surface to the photon detector area  
above and below the 
radiator vessel (see Fig.~\ref{hbdet}). 
The 116 mirror segments are mounted on rigid, low mass support structures 
inside the radiator volume and can be individually adjusted by stepper 
motors from the outside. 

All mirrors were first aligned after installation by surveying them inside the vessel.  
During the data taking periods, 
the mirror system was calibrated by making use of recorded events.
By comparing the charged particle track direction, obtained from the \v Cerenkov rings
due to  a particular spherical-planar mirror pair, to the track direction obtained
from other detectors of HERA-B, the calibration parameters of individual mirror segments,
as well as those of the entire RICH counter, could be extracted.
For this purpose, various data sets have been used. With the magnetic field turned off,
the direction of the straight tracks was accurately given by the target wire position
and the centroid of the cluster in the electromagnetic calorimeter. With the magnetic field turned on,
the tracks were determined by the tracking system. In order to reduce the uncertainty
in track direction due to multiple Coulomb scattering, only those tracks belonging to
particles with energy above 5 GeV were used.

\subsection{The calibration method}  
 
\begin{figure}[bht] 
\centerline{\epsfig{file=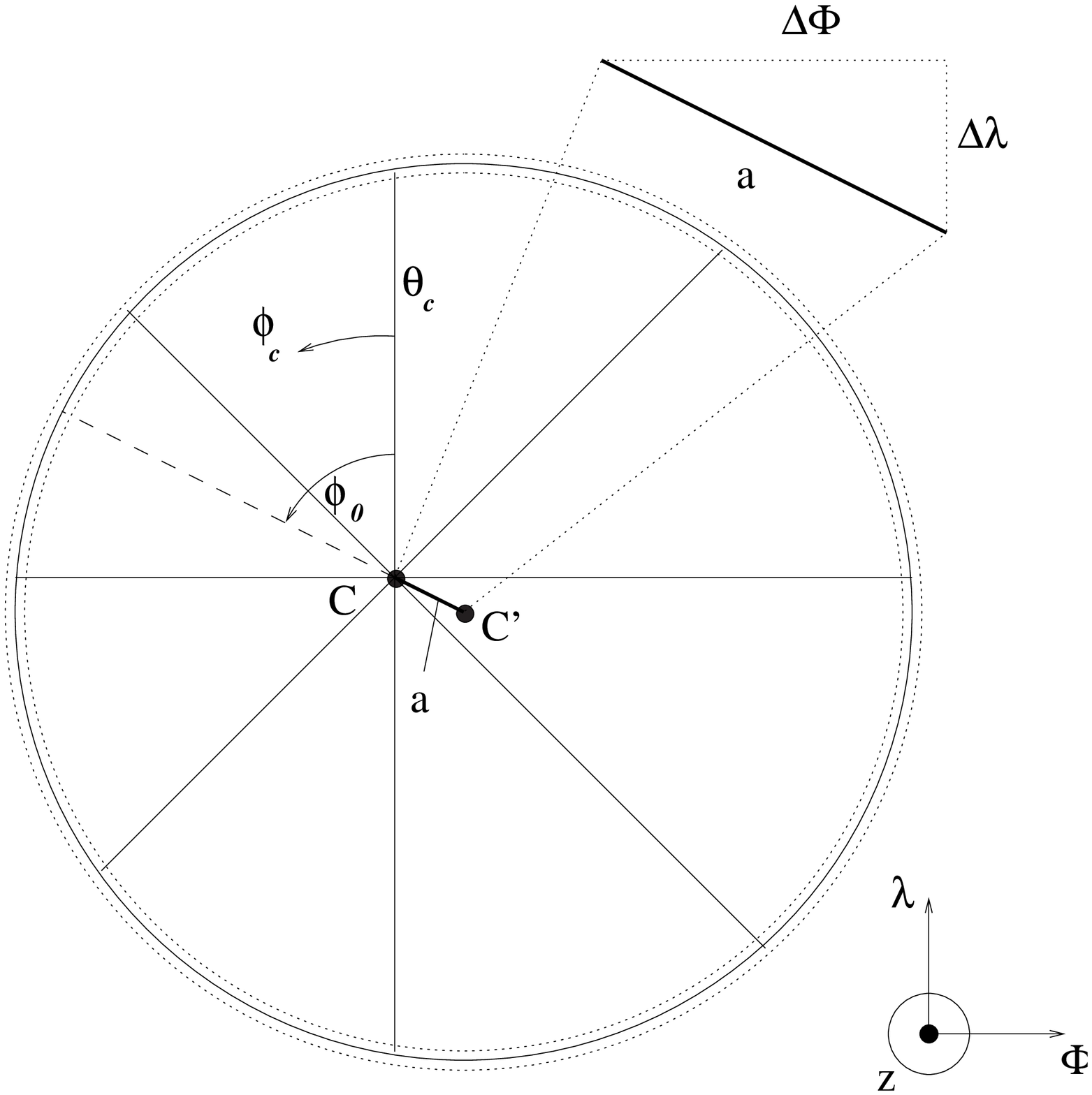,width=5cm,angle=-90.0} 
\epsfig{file=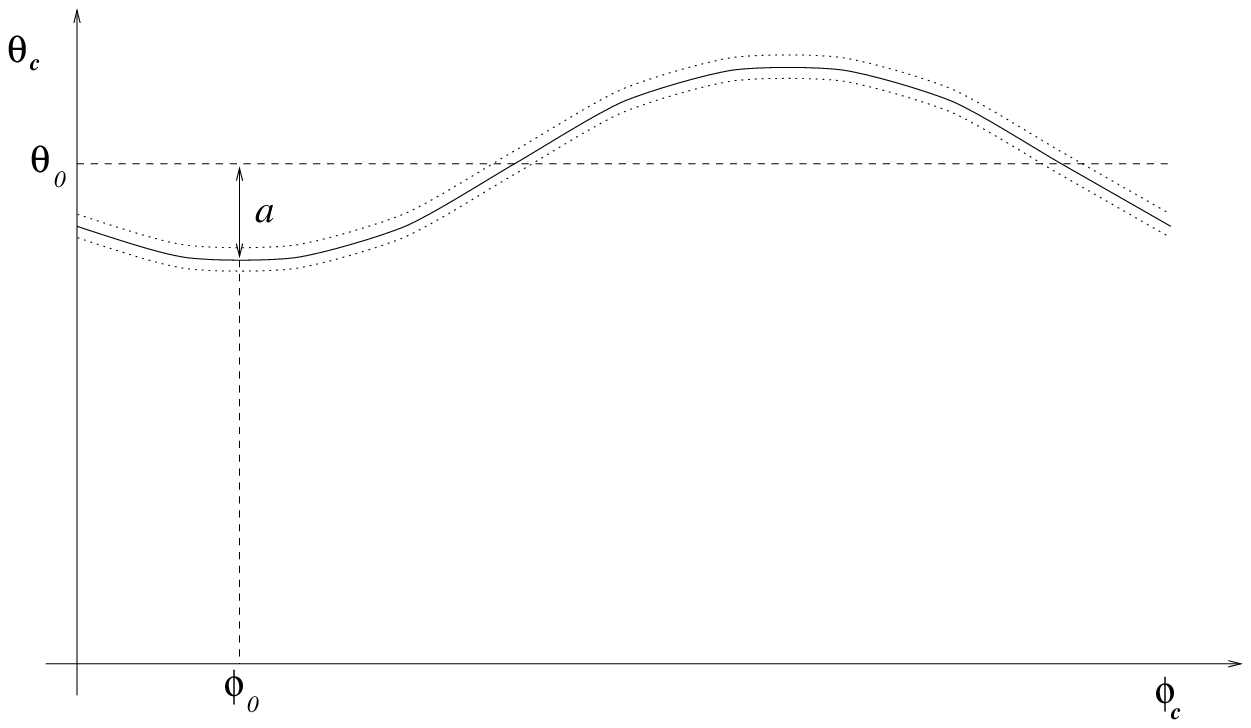,width=5cm,angle=-90.}} 
\caption{  For misaligned mirrors, the center of the measured ring C' is displaced
relative to the  extrapolated charged particle track direction  C (left). In such a case, the measured
\v Cerenkov angle $\theta _c$ depends on the azimuth $\phi _c$ of the photon hit (right).} 
\label{meth-scheme} 
\end{figure}

\begin{figure}[bht] 
\centerline{\epsfig{file=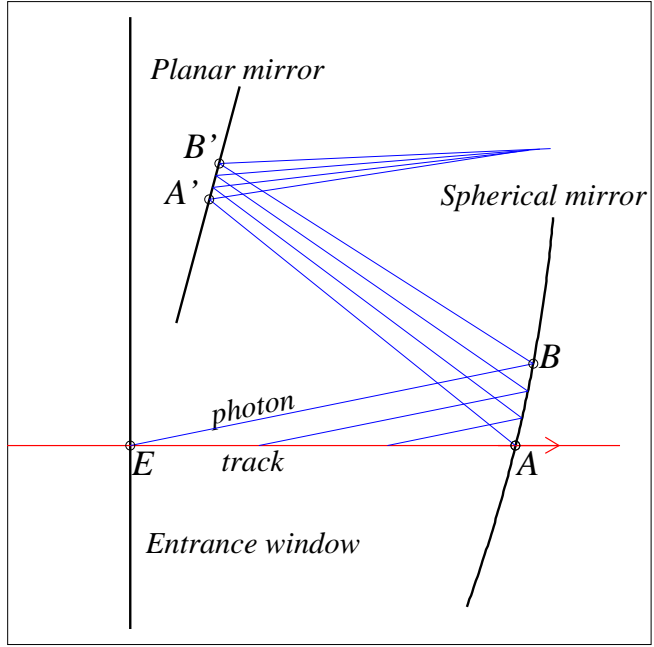,width=9cm}} 
\caption{   For calibration of mirror orientations, only those photon hits have been used, for which
the corresponding photons could have been reflected only from a given spherical-planar mirror pair,
regardless of the point on the particle track, from which the photon was emitted.
} 
\label{align_unamb} 
\end{figure}

Assume that one or both mirrors in a particular spherical-planar mirror pair are not well aligned.
In such a case, the measured \v Cerenkov ring, due to photons reflected on that pair,
will be displaced relative to the direction of the charged particle, which is taken to be reflected
on ideal mirror positions. For small displacements $a$, the azimuthal dependence of \v Cerenkov angle
for photons on a given ring is parameterized as (Fig.~\ref{meth-scheme}):
\begin{equation}\theta_{c} = \theta_0 +  a \cos (\phi _c - \phi _0)= \theta_0 + \Delta \Phi \cos{\phi_c} + 
\label{eq1}
 \Delta \lambda \sin{\phi_c},\end{equation} 
where $\theta_0$ is the nominal value of the \v Cerenkov angle.
The parameters $\Delta \Phi = a \cos \phi _0$ and $\Delta \lambda = a \sin \phi _0$
roughly correspond to rotations of the mirrors
around vertical and horizontal axes. A rotation of a spherical mirror segment 
by $\delta$ around the vertical and horizontal axes results in 
$\Delta \Phi = \delta$ and $\Delta \lambda = \delta$,
respectively, while the same rotation of the planar mirror results in 
$\Delta \Phi \approx \delta/2$ and $\Delta \lambda \approx \delta/2$
(see Fig.~\ref{align_unamb}).

The method obviously relies on accumulating a sufficient number of photons,
which have been reflected on a particular spherical-planar mirror pair. 
For each track-photon pair, the photon has been traced from two points on the
charged particle track to the photon hit position (Fig.~\ref{align_unamb}).
The first point is the particle entry into the radiator, the second is directly in front of
the spherical mirror. If both rays at a given azimuthal angle are reflected from the 
same spherical-planar mirror
pair, such a photon hit is taken as a valid data point on the \v Cerenkov ring, relevant
for that mirror pair. In other words, a photon hit is valid for our analysis
if the corresponding photon would have been reflected from the same spherical-planar mirror pair, regardless
of the point on the charged particle trajectory from which it might have originated.
In the following we refer to such hits as calibration hits. 

\begin{figure}[bht] 
\centerline{\psfig{file=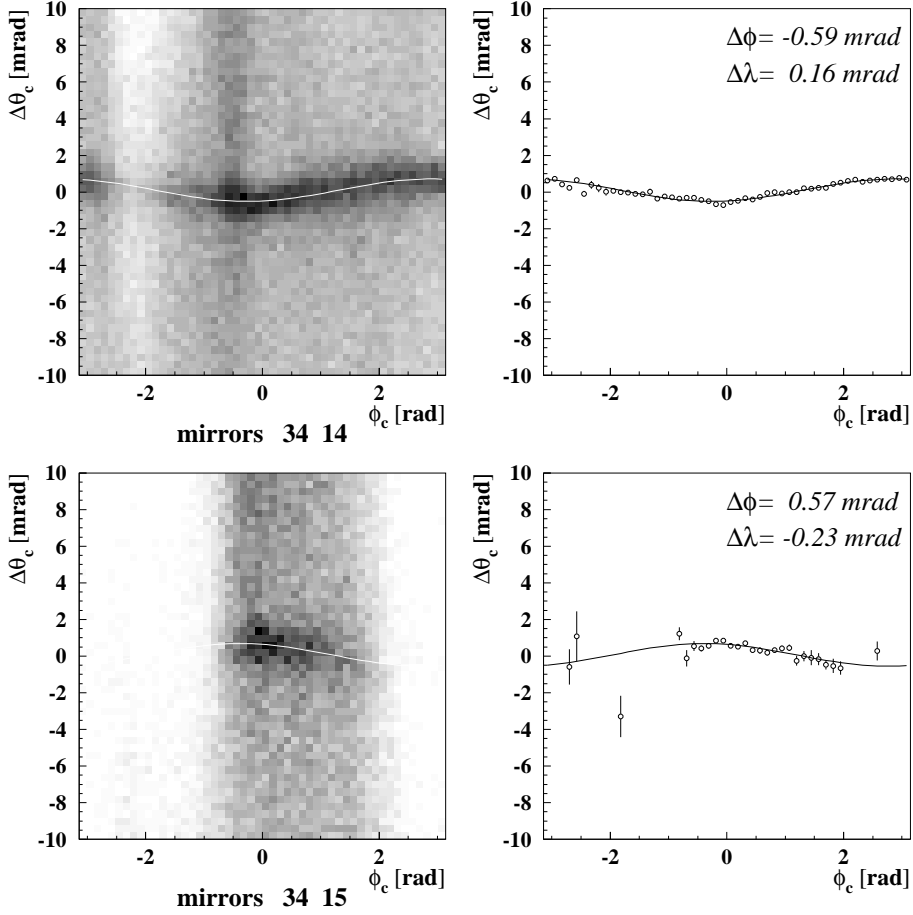,width=13cm}} 
\caption{  Distribution of hits in the $\Delta \theta_{c}$,  $\phi_c$ plane for  
two combinations of a spherical and two planar mirror segments. Raw data 
are shown on the left side, and the  
ring peak position from the fit in each 
 $\phi_c$ slice is shown on the right. } 
\label{align_thc_vs_phi_data} 
\end{figure}

\begin{figure}[bht] 
\centerline{\psfig{file=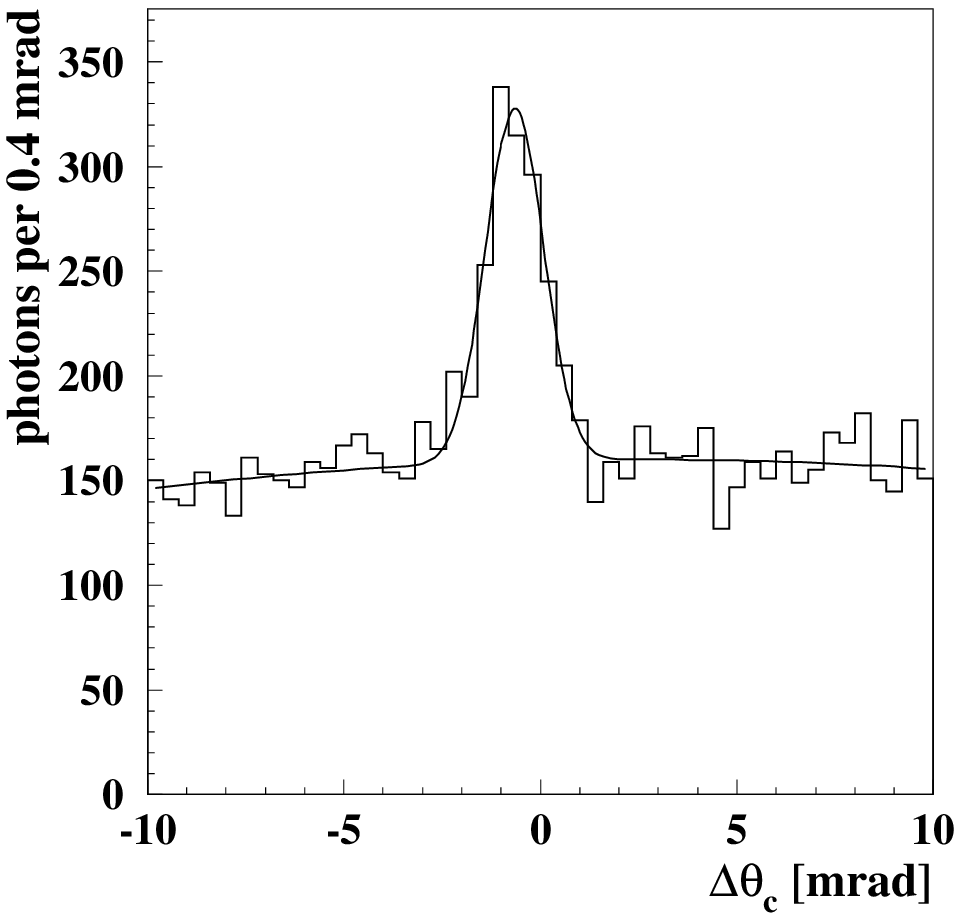,width=10cm}} 
\caption{    
Distribution of hits in $\Delta \theta_{c}$ for one of the  $\phi_c$  
slices. The result of the fit is superimposed. 
} 
\label{align_fit_slice} 
\end{figure}

Figure~\ref{align_thc_vs_phi_data}  shows the distribution of 
such hits for two particular combinations of a spherical and two planar  mirrors.
The two-dimensional 
histograms on the left hand side of the figure represent the number of calibration hits
as a function of 
\v Cerenkov angle difference $\Delta \theta_c = \theta_c - \theta_c^{\pi}(p)$ and azimuthal angle 
$\phi_c$. The difference to the nominal \v Cerenkov angle $\theta_{c}^{\pi}(p)$ of a pion at the 
measured momentum $p$ is chosen in order to suppress the momentum dependence; the pion
hypothesis is chosen since the 
majority of tracks correspond to pions. An accumulation of hits at about $\Delta \theta_c = 0$ 
can be seen, with some modulation dependent on azimuthal angle $\phi _c$.
For each of 50 slices in $\phi _c$, the distribution was fitted with a Gaussian for the peak and
a polynomial background (Fig.~\ref{align_fit_slice}). The \v Cerenkov peak position as a function of 
azimuthal angle (right hand side of Fig.~\ref{align_thc_vs_phi_data}) 
is then fitted with the function $\Delta \Phi \cos{\phi_c} + \Delta \lambda \sin{\phi_c} + C$,
and the rotation angles $\Delta \Phi$ and $\Delta \lambda$ are obtained for the given 
mirror pair\footnote{The parameter 
$C$ is a constant to account for a possible bias in the \v Cerenkov angle measurement.}.

Close to mirror boundaries, parts of a ring could be shared by different combinations of spherical
and planar mirrors, 
which permits relative calibration of adjacent mirrors.

\subsection{Extraction of calibration parameters of individual mirror segments}

The measured displacement $(\Delta \Phi, \Delta \lambda$) 
for a pair of mirrors is equal to the sum of unknown  contributions
from the corresponding spherical and planar mirror segments $k$ and $j$:
\begin{equation} 
\label{eq2}
\Delta \Phi_{kj} = \Delta \Phi^{sph}_k +  
\Delta \Phi^{pl}_j,\end{equation} 
\begin{equation} 
\label{eq3}
\Delta \lambda_{kj} = \Delta \lambda^{sph}_k +  
\Delta \lambda^{pl}_j.\end{equation} 
This represents a system of $2m$ linear  equations for the $2n$ unknown parameters, where $m$ is the number
of all measured segment combinations $(k,j)$, and $n$ is the number of mirror segments.  
As any given mirror segment
contributes to more than one combination, there are more equations than there are unknowns.
Arranging the measured $\Delta \Phi _{kj}$ and $\Delta \lambda _{kj}$ into vectors $b_i^{\Phi}$ 
and $b_i^{\lambda}$, with dimensions equal to $m$, 
the system of equations may be written for each of the detector halves as
\begin{equation} 
\label{eq4}
b_i^{\Phi} = \sum_{l=1}^{n} A_{il} u_l^{\Phi},  \ \ \ \  
b_i^{\lambda} = \sum_{l=1}^{n} A_{il} u_l^{\lambda}, \ \ \ \ i=1,...,m, \end{equation} 
where $u_l^{\Phi}$ and $u_l^{\lambda}$ are the unknown contributions to the rotation
$(\Delta \Phi, \Delta \lambda$) of a particular mirror segment. 
Note that the matrix $A_{il}$ is of a particular simple form; in a given row it only has 1
at two places (given by the indices of the spherical and planar segments, Eqs.~\ref{eq2} 
and \ref{eq3}), and is otherwise equal to zero.

Note that we have to add an additional condition to fully determine the 
system: if all spherical mirrors are turned by $\delta$, and all planar  
mirrors by $\approx - 2 \delta$, one arrives at the same ring displacements  
(vectors $b_i^{\Phi}$ and $b_i^{\lambda}$). Rather than fixing the displacement of one  
mirror in the lower and one in the upper half to zero (one of the  
possibilities), we require of the solution to  
minimize the necessary mirror readjustments, 
\begin{equation} 
\label{eq5}
\sum_{i_s}  u_{i_s}^{\Phi} -  \sum_{i_p} u_{i_p}^{\Phi} = 0, \ \ \ \ 
\sum_{i_s}  u_{i_s}^{\lambda} -  \sum_{i_p} u_{i_p}^{\lambda} = 0.\end{equation} 
The first sum in each of the equations runs over all spherical mirror segments and the second over 
all planar ones in a given detector half. We include these two equations 
in the system of equations \ref{eq4}, and get two systems with $m+1$ equations,
\begin{equation} 
\label{eq5.1}
b_i^{\Phi} = \sum_{l=1}^{n} A'_{il} u_l^{\Phi},  \ \ \ \  
b_i^{\lambda} = \sum_{l=1}^{n} A'_{il} u_l^{\lambda}, \ \ \ \ i=1,...,m+1, \end{equation} 
where $A'$ replaces $A$ to account for the additional equation.

The systems of equations, Eq.~\ref{eq5.1}, are  
solved by requiring that the properly weighted sum of squares of 
deviations  of the left hand side from the right hand side for each of the two systems 
 is minimal,  
\begin{equation} 
\sum_{i=1}^{m+1} \frac{(\sum_{l=1}^{n} A'_{il} u_l -  
b_i)^2}{\sigma_i^2} = min,\end{equation} 
where the terms are weighted by the inverse square of the error  
$\sigma_i$  in the measurement\footnote{For the two additional equations \ref{eq5}, 
a value of $\sigma_i=0.01$~mrad was assumed; no influence was found when this value was varied.}
 of $b_i$. Here $u_l$ and $b_i$ denote either  $u_l^{\Phi}$ and $b_i^{\Phi}$
or $u_l^{\lambda}$ and $b_i^{\lambda}$.
The resulting linear system, 
\begin{equation}   
\sum_{l=1}^{n} \sum_{i=1}^{m+1} \frac{ A'_{il} A'_{ij}}{\sigma_i^2} u_l = 
 \sum_{i=1}^{m+1} \frac{A'_{ij}}{\sigma_i^2} b_i \end{equation} 
 is readily solved, 
\begin{equation}u_k = \sum_{j=1}^{n}  (B^{-1})_{kj} \sum_{i=1}^{m+1}  
\frac{A'_{ij}}{\sigma_i^2} b_i.\end{equation} 
Here we have defined
a new matrix \begin{equation}B_{lj}=\sum_{i=1}^{m+1} \frac{A'_{il} A'_{ij}}{\sigma_i^2}, 
\end{equation} 
which is symmetric, so that its inverse ${\bf B^{-1}}$ 
is easy to calculate. 
The resulting errors on $u_k$ are given by
\begin{equation}\sigma_{u_k}^2 = \sum_{i=1}^{m+1} \bigl( \sum_{j=1}^{n}  (B^{-1})_{kj} 
\frac{A'_{ij}}{\sigma_i}  \bigr)^2.\end{equation}

\subsection{Calibration results}  
 
\begin{figure}[bht] 
\centerline{\psfig{file=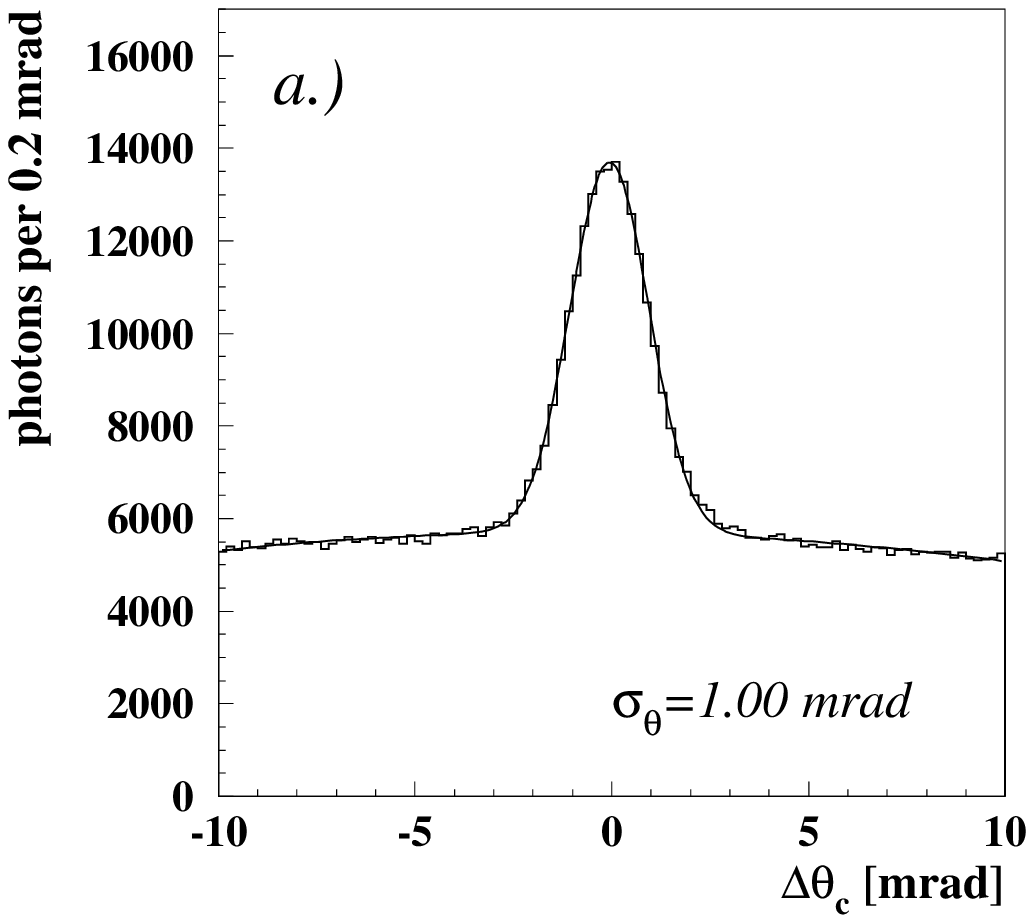,width=7cm}\psfig{file=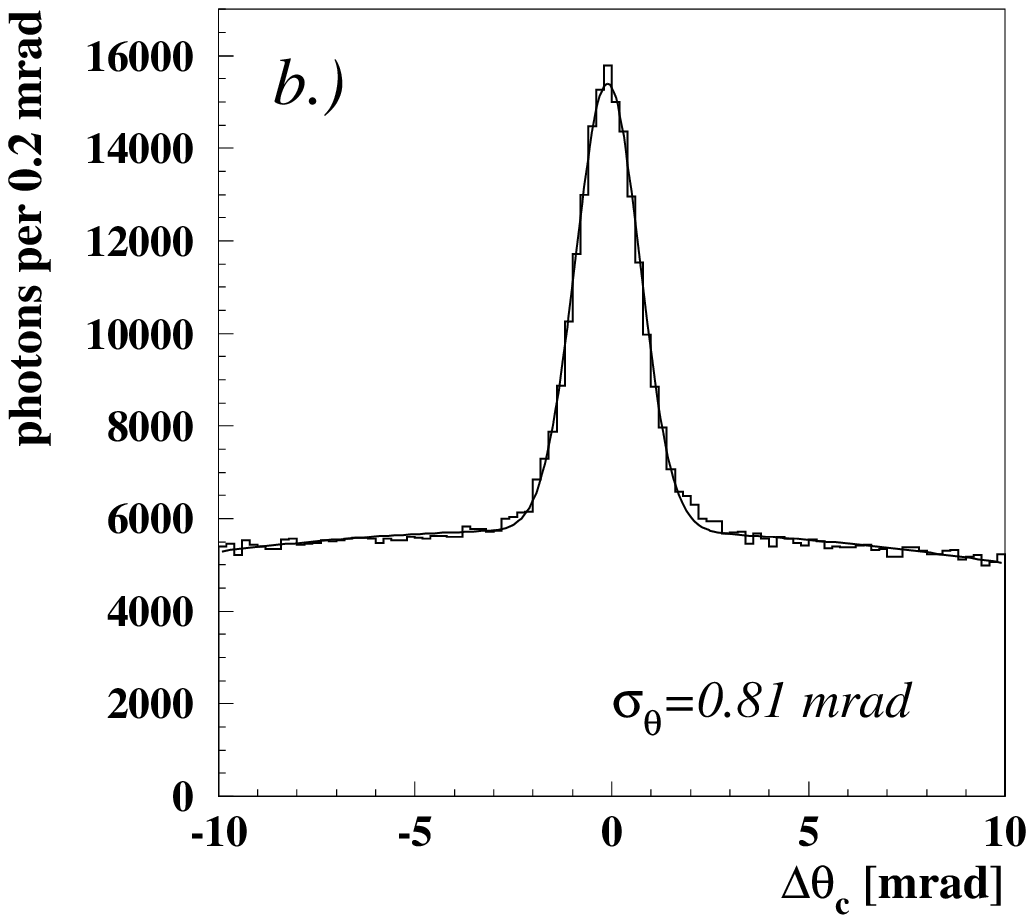,width=7cm}} 
\caption{    
The distribution of photon hits with respect to the corresponding \v Cerenkov angle difference
$\Delta \theta_c$
before (a) and after (b) applying the correction obtained from the calibration of mirrors.
} 
\label{align_thc_data} 
\end{figure}

Having solved the system of equations for angular displacements of  
individual mirror segments, the quality of the new alignment is  checked on the data,
by examining the \v Cerenkov angle resolution before and after applying the new 
calibration parameters, 
$ \Delta \Phi^{sph}_k$,  
$ \Delta \lambda^{sph}_k$, 
$ \Delta \Phi^{pl}_j$, and 
$ \Delta \lambda^{pl}_j$. 
 The result is shown in Fig.~\ref{align_thc_data}; 
the improvement in resolution is clearly visible.

To check the resulting alignment of the optical system, we have reanalyzed the data    
after having applied the calculated corrections. As expected, the new corrections were
consistent with zero.

To study possible systematic effects,  
we have investigated different data sets, recorded under different 
conditions. We found  good agreement of the results  
with two different gas radiators, when freon ($\theta _c$=52 mrad) was replaced with nitrogen 
($\theta _c$=25 mrad). 
A similar analysis was also performed on the data recorded without magnetic  
field. Again the analysis yielded results that are  
consistent with the values deduced from the data with magnetic field. 
We also found that  the alignment
parameters did not change over extended periods of time,
which testifies to stable mirror positions.

Possible systematic effects were also checked by using Monte Carlo generated events, 
where all mirror segments were assumed to be  
perfectly aligned. Although the resulting parameters are all  
consistent with zero, small systematic effects at the level of 0.1~mrad could not be excluded
due to limited statistics.

\section{Conclusions} 
 
A method was developed for determination of the 
alignment of the RICH counter relative to other parts of the spectrometer,  
either the system of tracking chambers or the electromagnetic calorimeter. 
The method was tested on various sets of real data, 
recorded with and without magnetic field, as well as 
 on simulated data.  
By applying alignment corrections for each mirror  
segment as derived by this method, a significant improvement in the resolution 
of the \v Cerenkov angle measurement could be obtained.

The HERA-B experiment finished data taking in spring 2003. During its five years  
of operation, the HERA-B RICH has proved reliability and stability of all its components 
and especially of the multi anode PMTs. 
The low noise, high rate capability and excellent long term stability of these devices 
enabled excellent operation in the hostile environment of a hadron machine. 
The HERA-B RICH identifies pions, kaons and protons essentially in the entire  
kinematic range of the HERA-B experiment  
with the identification efficiencies as large as 90\% 
and mis-identification probability at the 1\% 
level~\cite{rich_nim,staric-rich04}. With kaon and proton  
identification the combinatorial background is in some cases reduced by more than  
3 orders of magnitude. Several physics analyses
would not have been possible  
without the excellent performance of the RICH particle identification 
system~\cite{staric-rich04,PQ,kstar-phi,open-charm}.

\end{document}